\documentclass[conference]{IEEEtran}
\IEEEoverridecommandlockouts

\usepackage{cite}
\usepackage{amsmath,amssymb,amsfonts}
\usepackage{algorithmic}
\usepackage{graphicx}
\usepackage{textcomp}
\usepackage{xcolor}
\usepackage{booktabs}
\usepackage{multirow,multicol}
\usepackage{float}
\usepackage{tabularx}
\usepackage{subfigure,subcaption,subfig}
\usepackage[inkscapelatex=false]{svg}
\def\BibTeX{{\rm B\kern-.05em{\sc i\kern-.025em b}\kern-.08em
    T\kern-.1667em\lower.7ex\hbox{E}\kern-.125emX}}

\begin{document}

\bstctlcite{IEEEexample:BSTcontrol}

\title{Characterizing Communication Patterns in Distributed Large Language Model Inference
}

\makeatletter
\newcommand{\linebreakand}{%
  \end{@IEEEauthorhalign}
  \hfill\mbox{}\par
  \mbox{}\hfill\begin{@IEEEauthorhalign}
}
\makeatother

\author{
  \IEEEauthorblockN{Lang Xu}
  \IEEEauthorblockA{\textit{Dept. Computer Science \& Engineering} \\
    \textit{The Ohio State University}\\
    Columbus, OH, United States \\
    xu.3304@osu.edu}
  \and
  \IEEEauthorblockN{Kaushik Kandadi Suresh}
  \IEEEauthorblockA{\textit{Dept. Computer Science \& Engineering} \\
    \textit{The Ohio State University}\\
    Columbus, OH, United States \\
    kandadisuresh.1@osu.edu}
  \and
  \IEEEauthorblockN{Quentin Anthony}
  \IEEEauthorblockA{\textit{Dept. Computer Science \& Engineering} \\
    \textit{The Ohio State University}\\
    Columbus, OH, United States \\
    anthony.301@osu.edu}
  \linebreakand 
  \IEEEauthorblockN{Nawras Alnaasan}
  \IEEEauthorblockA{\textit{Dept. Computer Science \& Engineering} \\
    \textit{The Ohio State University}\\
    Columbus, OH, United States \\
    alnaasan.1@osu.edu}
  \and
  \IEEEauthorblockN{Dhabaleswar K. Panda}
  \IEEEauthorblockA{\textit{Dept. Computer Science \& Engineering} \\
    \textit{The Ohio State University}\\
    Columbus, OH, United States \\
    panda@cse.ohio-state.edu}
}

\maketitle

\begin{abstract}
Large Language Models (LLMs) built on transformer architectures have transformed natural language processing, achieving remarkable performance across diverse applications. While distributed inference frameworks enable practical deployment of these models, inter-GPU communication creates significant performance constraints that limit service quality in real-world systems. This paper investigates communication dynamics in distributed LLM serving—analyzing how various parallelization approaches coordinate data exchange between GPU workers during inference. We study dense transformer-based models as representative examples of contemporary architectures widely used in operational deployments. Our work combines detailed profiling measurements with predictive analytical models to characterize communication behavior across different parallelization configurations. Results show that tensor parallelism incurs substantial network overhead but delivers superior response times for brief sequences, pipeline parallelism minimizes data transfer requirements while increasing total latency, and combined approaches demand careful tuning to achieve balanced performance. These insights offer practical recommendations for selecting appropriate parallelization schemes in production LLM services and identify key opportunities for optimizing inference frameworks and communication infrastructure.
\end{abstract}


\section{Introduction}
\label{sec:intro}

Large Language Models (LLMs) have been demonstrating exceptional capabilities across multiple modalities, including natural language understanding, vision, and speech processing. Exemplified by models such as Llama 3 \cite{grattafiori2024llama3herdmodels}, Claude 3 \cite{TheC3}, and GPT-4 \cite{openai2024gpt4}, these foundation models undergo extensive pre-training to develop sophisticated understanding of human language and reasoning patterns. Recently, post-training techniques such as reinforcement learning \cite{ouyang2022traininglanguagemodelsfollow} and test-time scaling \cite{zhang2025surveytesttimescalinglarge} have enabled the emergence of specialized models like DeepSeek-R1 \cite{deepseekai2025deepseekr1incentivizingreasoningcapability}, OpenAI o1 \cite{openai2024openaio1card}, and Gemini Pro \cite{blogGemini25} that excel in complex reasoning, detailed explanation, and strategic planning \cite{wei2023chainofthoughtpromptingelicitsreasoning, zhang2025surveytesttimescalinglarge}. Unlike pre-training, which primarily scales with model parameters and training data volume \cite{10664295}, post-training techniques leverage extended inference-time computation to elicit more deliberate reasoning and enhanced in-context learning capabilities \cite{zhang2025surveytesttimescalinglarge}.

\subsection{Motivation}
\label{sec:motivation}

LLM inference shares fundamental characteristics with training in its requirement for distributed computing resources: multiple GPUs are needed to host large model parameters, maintain extensive key-value caches \cite{pope2022efficientlyscalingtransformerinference}, and enable efficient parallel computation. However, both paradigms are constrained by the same critical bottleneck—inter-GPU communication required to synchronize workers and maintain computational correctness. During inference, the autoregressive token generation process following the initial prefill stage \cite{zhong2024distservedisaggregatingprefilldecoding} creates unique communication patterns distinct from training, where the sequential nature of decoding and limited computation-communication overlap means that communication overhead constitutes a larger proportion of total execution time (Figure~\ref{fig:comm_comp}). Despite substantial research on distributed training communication, there remains a significant gap in understanding the detailed communication characteristics of multi-GPU inference workloads and how different parallelism strategies impact end-user experience metrics and service level objectives (SLOs). Furthermore, practitioners currently lack systematic guidance on when and how to select appropriate parallelism configurations for specific inference scenarios, creating a critical knowledge gap that hinders optimal deployment decisions in production environments where communication efficiency directly impacts user experience and service quality.

\begin{figure}[htbp]
    \centering
    \includegraphics[width=\columnwidth]{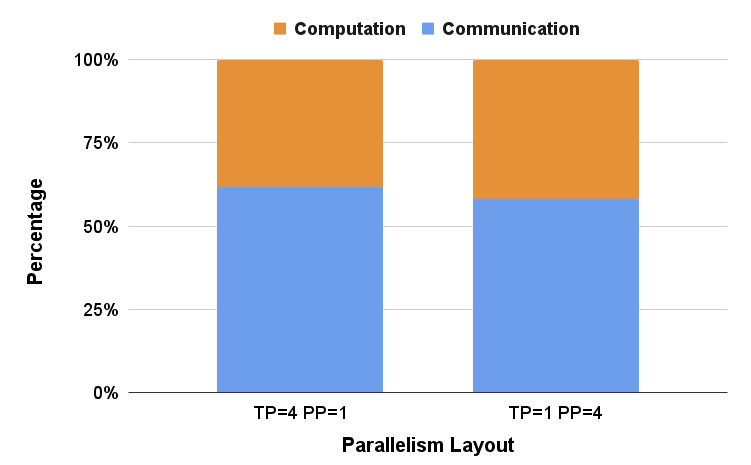}
    \caption{Communication-computation breakdown for Llama-3.1-8B inference under various parallelism settings}
    \label{fig:comm_comp}
\end{figure}

\subsection{Problem Statement}
\label{sec:problem-stmt}

This paper systematically characterizes the communication patterns occurring within state-of-the-art high-performance LLM inference frameworks deployed on GPU clusters. Our objective is to analyze GPU communication patterns and protocols, including collective operations and point-to-point communications, across different parallelism strategies including tensor parallelism, pipeline parallelism, and hybrid parallelism approaches. Our analysis encompasses comprehensive examination of communication volumes, operation frequencies, and message size distributions under various parallelism schemes and worker counts. Furthermore, we examine how SLO performance varies across different parallelism configurations when serving parameters and model architectures are modified, providing insights into optimal parallelism selection for diverse deployment scenarios.

\subsection{Challenges}
\label{sec:challenges}

Building upon the initial motivation discussed, we aim to address the following overarching challenges:

\begin{itemize}
\item What are the predominant types, volumes, and patterns of communication occurring during distributed LLM inference when scaled across multiple GPU workers using state-of-the-art model inference frameworks?

\item Can we develop analytical models to predict communication characteristics given specific inference configurations, including parallelism degree, model architecture, and serving parameters?

\item What is the impact of communication volume and patterns on user experience metrics and standard LLM inference SLOs, such as end-to-end inference latency, time-to-first-token and time-per-output-token?

\item What is the comparative impact of different parallelism layouts on communication overhead when hosting models, and what optimization insights can be derived from this analysis?
\end{itemize}

\subsection{Proposed Solution}
\label{sec:proposed-solution}

Given the complexity of understanding communication in distributed LLM inference, we adopt a systematic methodology that combines empirical profiling with analytical modeling to study communication patterns across various parallelism strategies, model architectures, and serving scenarios. Our approach aims to provide comprehensive understanding of communication overheads in LLM deployment using widely-adopted high-performance inference frameworks, specifically vLLM\cite{kwon2023efficientmemorymanagementlarge}.

Our characterization encompasses multiple parallelism strategies including Tensor Parallelism, Pipeline Parallelism, and hybrid approaches for models up to 13B parameters. We conduct detailed profiling throughout the inference pipeline to measure communication patterns for each parallelism strategy, capturing: 1) communication primitive types, 2) network data volumes, 3) operation frequencies and message size distributions for each communication type. To align with service-level requirements, we scale our analysis across varying sequence-lengths, studying their impact on resulting communication volumes. Finally, we conduct multi-node scaling studies with various parallelism configurations to derive insights for optimal parallelism selection in production inference deployments.

\subsection{Contributions}
\label{sec:contribution}

Our contributions are as follows:

\begin{enumerate}
    \item We present the first systematic study of communication behavior in distributed LLM inference, developing analytical models that predict communication patterns across various model sizes, inference stages, serving scenarios, and parallelism strategies.

    \item We develop and validate analytical models for estimating communication volume for Tensor Parallelism, Pipeline Parallelism, and hybrid schemes across different sequence lengths and model architectures.

    \item We deliver extensive empirical measurements of communication statistics for each parallelism configuration, including collective operation types, network data volumes, message size distributions, and operation frequencies.

    \item We examine the impact of different parallelism schemes on critical inference metrics including time-to-first-token, token generation throughput, and per-token latency, providing actionable insights for production deployments.
\end{enumerate}

\subsection{Paper Breakdown}
\label{sec:breakdown}

The rest of the paper follows this structure. Section \ref{sec:background} provides the necessary background on LLM inference, deployment strategies, serving metrics, as well as various parallelism schemes used to host models on large-scale distributed systems. Section \ref{sec:analysis} details our analytical models for predicting communication volume across different parallelism schemes and derives the theoretical foundations for our analysis. Section \ref{sec:results} presents our experimental results and profiling data, validating our analytical models against empirical measurements. Section \ref{sec:related} reviews related work in LLM inference optimization and characterization. Section \ref{sec:conclusions} concludes this work with key insights and recommendations for future research directions.

\section{Background}
\label{sec:background}
\vspace{-1ex}
Efficiently serving Large Language Models (LLMs) for inference is critical as their applications expand. This section covers the fundamentals of LLM inference and the parallelism schemes that are vital for their deployment.
\begin{figure}[htbp]
\centering
\includegraphics[width=\columnwidth]{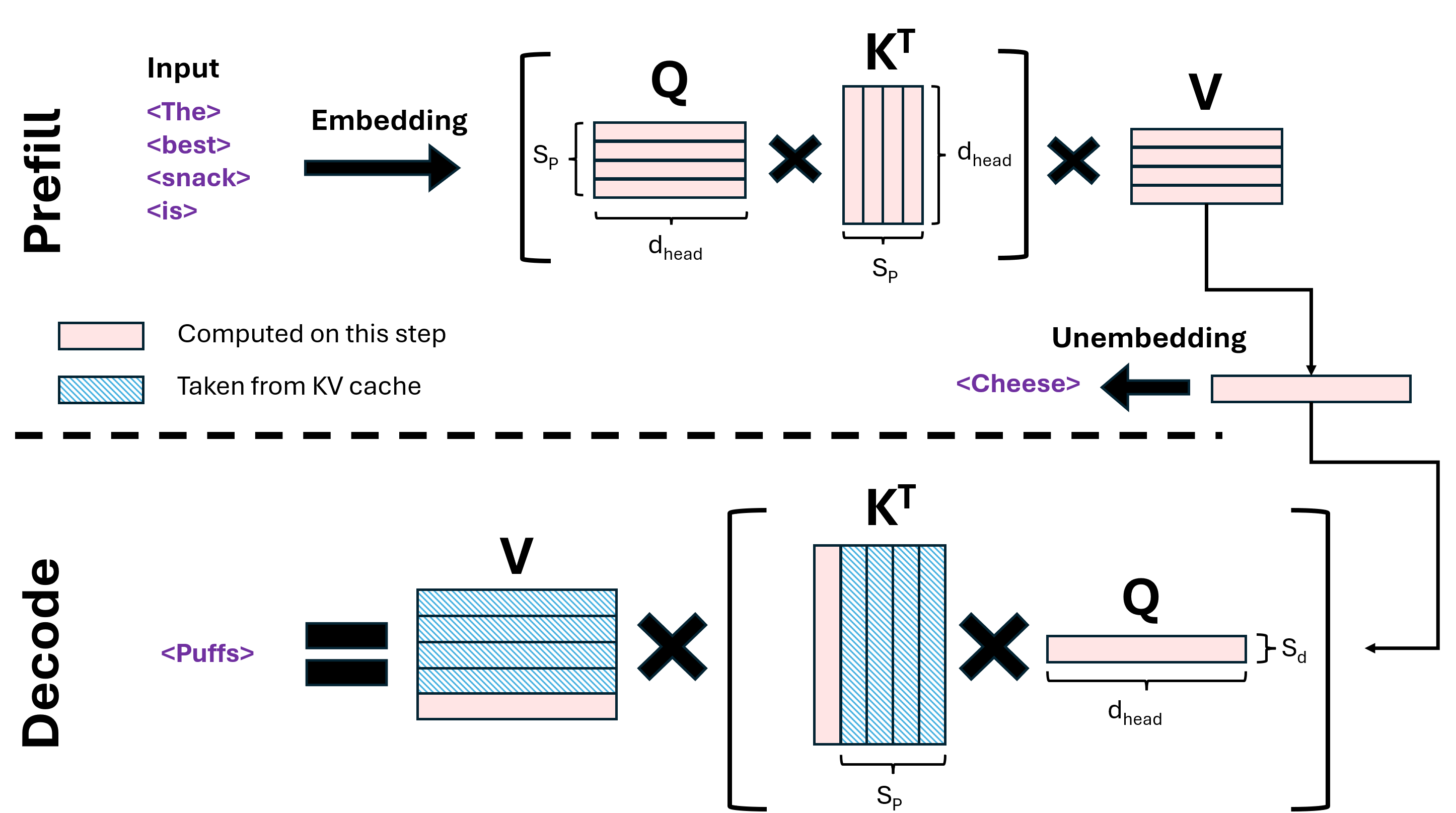}
\caption{Illustration of general LLM inference}
\label{fig:inference-workflow}
\end{figure}
\vspace{-0.5ex}
\subsection{Large Language Model Inference}
\vspace{-0.5ex}
LLM inference is typically an autoregressive process that involves two distinct phases. The prefill phase processes the input prompt in parallel, computing initial Key-Value (KV) states for attention that are stored in a KV cache. This is followed by a sequential decoding phase, where tokens are generated one by one, with each step utilizing the KV cache (Figure~\ref{fig:inference-workflow}).
Key performance metrics include Time-To-First-Token (TTFT), which measures the latency from when a request is received to when the first output token is generated, and Time-Per-Output-Token (TPOT), which represents the average time taken to generate each subsequent token after the first. Overall throughput is measured as the number of output tokens generated per second or requests processed per second by the system.
The increasing importance of LLM inference is driven by their expanding capabilities. Modern LLMs perform not only language completion but also complex reasoning, code generation, and interaction with external tools. Techniques like test-time scaling and advanced prompting strategies further enhance these abilities during inference, often requiring more extensive computation per request than simple generation tasks. This necessitates highly optimized inference frameworks and a deep understanding of performance bottlenecks, including communication overhead.
\vspace{-0.5ex}
\subsection{Parallelism Schemes for Inference}
\vspace{-0.5ex}
LLMs often exceed single-GPU memory capacity and performance targets, necessitating parallelism. Inference parallelism differs from training parallelism in that it involves only a forward pass (no gradients), and managing the KV cache is a primary concern. The focus is often on achieving low latency for individual requests or maximizing throughput of concurrent requests.
Tensor Parallelism (TP), pioneered by Megatron-LM \cite{shoeybi2019megatron}, implements intra-layer parallelism by partitioning weight matrices of operators like GEMMs within transformer layers (e.g., in MLPs and attention blocks) across multiple GPUs. This involves splitting matrices column-wise or row-wise. For example, a column-parallel GEMM followed by a row-parallel GEMM requires an All-Reduce collective operation to sum partial results before proceeding. TP is communication-intensive, relying on high-bandwidth intra-node interconnects. As detailed in Section~\ref{sec:analysis}, this typically results in specific collective calls per layer (e.g., All-Reduce, Gather).
Pipeline Parallelism (PP) \cite{huang2019gpipe} implements inter-layer parallelism, distributing entire layers of the model across different GPUs to form pipeline stages. Activations are passed point-to-point between GPUs in adjacent stages. While conceptually simple, this can lead to "pipeline bubbles" or GPU idle times, especially with single requests. Communication involves send/recv operations for activations, with $p-1$ for $p$ stages.

Hybrid Parallelism combines TP and PP, often using TP within compute nodes and PP across nodes, to scale to larger models and GPU counts. This approach balances the communication demands of each strategy against the available interconnect capabilities. Detailed illustrations of these parallelism schemes are shown in Figure~\ref{fig:parallel}.
\begin{figure}[htbp]
\centering
\includegraphics[width=\columnwidth]{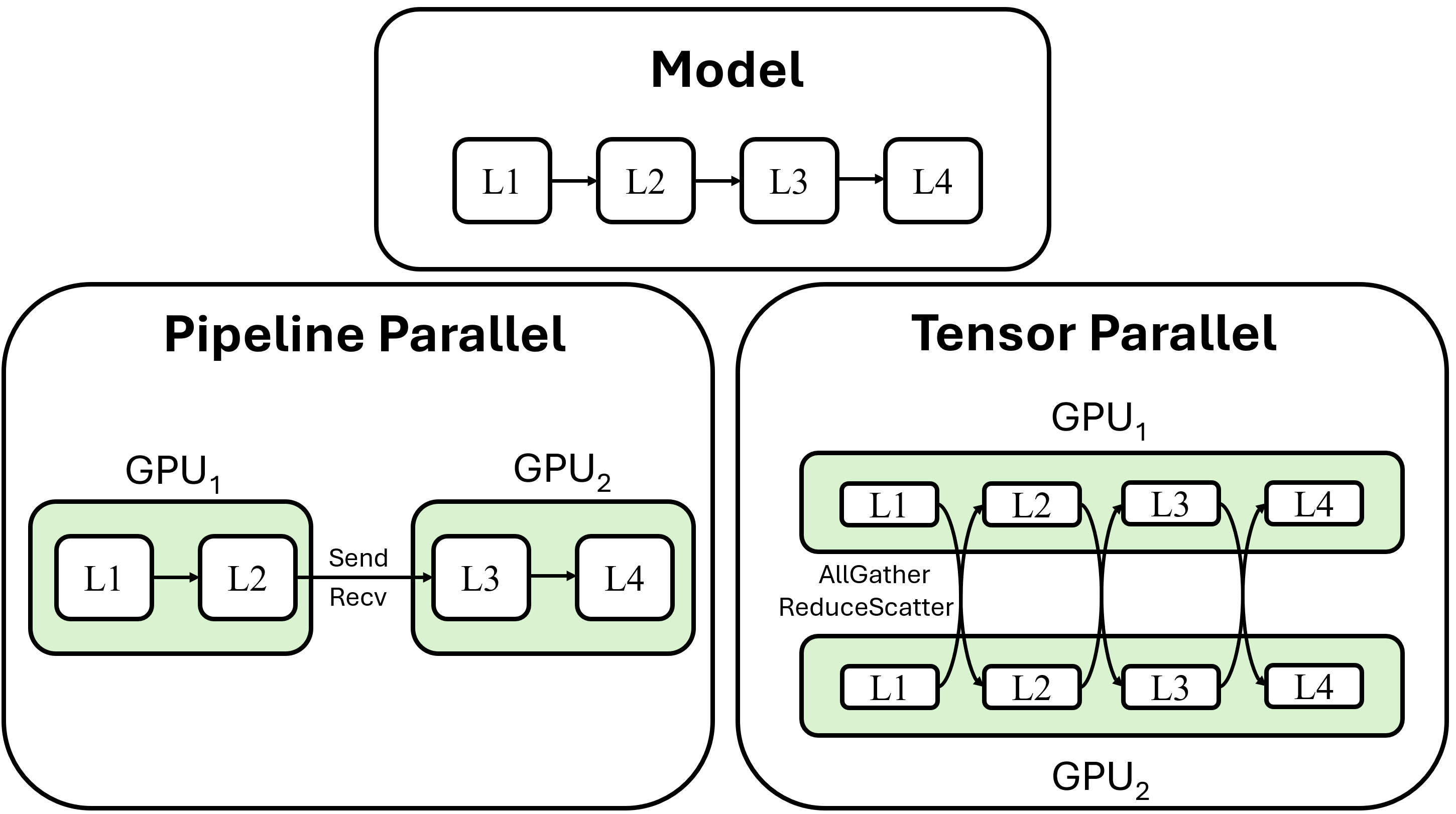}
\caption{Illustration of Tensor and Pipeline Parallelism}
\label{fig:parallel}
\end{figure}
\section{Analysis}
\label{sec:analysis}

\begin{table}[ht]
\resizebox{\columnwidth}{!}{%
\begin{tabular}{l|ll|l}
\toprule
\textit{h} & Hidden dimension size        & \textit{t} & Tensor-parallel size \\
\textit{L} & Number of transformer layers & \textit{p} & Pipeline-parallel size \\
\textit{b} & Bytes per element            & \textit{v} & Vocabulary size \\
\textit{$S_p$} & Prefill sequence-length    & \textit{$S_d$} & Decode sequence-length \\
\textit{$a$} & Number of attention heads    & \textit{$d_{head}$} & Head dimension \\
\bottomrule
\end{tabular}%
}
\vspace{2ex}
\caption{Variable definitions.}
\label{tab:varnames}
\end{table}
This section presents our analytical models for predicting communication volume across different parallelism strategies in distributed LLM inference. We systematically derive communication patterns for tensor parallelism, pipeline parallelism, and their hybrid combination, using the dense Llama transformer architecture implemented in vLLM as our reference framework.

\subsection{Tensor Parallelism}

Tensor parallelism distributes computation across multiple GPUs by partitioning matrix operations within individual transformer layers. In the prevalent implementation, linear layers employ row-parallel distribution where input matrices are partitioned along the second dimension and weight matrices along the first dimension. For a linear transformation $Y = XA + b$ distributed across $t$ tensor-parallel devices, each GPU computes partial results $Y_i = X_i A_i$ for $i \in \{0, ..., t-1\}$, requiring subsequent synchronization to materialize the complete output.

Row-parallel linear layers appear at two critical locations within each transformer block: (1) the MLP down-projection layer that reduces from the expanded intermediate dimension back to the hidden size $h$, and (2) the attention output projection that combines multi-head attention results. Each transformer layer therefore generates two Allreduce operations with message sizes of $h$ elements. Additionally, the embedding layer requires one Allreduce operation per decoded token, while logit computation necessitates a Gather operation across tensor-parallel workers.

For a complete inference request encompassing both prefill ($S_p$ tokens) and decode ($S_d$ tokens) phases, the total communication volume under pure tensor parallelism is:

\begin{equation}
V_{tp} = (2L + 1) \times (S_p + S_d - 1) \times h \times b \times 2\left(\frac{t-1}{t}\right) + S_d \times \frac{v}{t} \times b
\end{equation}

where the first term captures Allreduce operations with the standard correction factor $2(t-1)/t$, and the second term represents Gather operations for vocabulary projection.

\subsection{Pipeline Parallelism}

Pipeline parallelism partitions transformer layers across multiple devices, requiring point-to-point communication to transfer intermediate activations and key-value cache states between pipeline stages. During the prefill phase, each pipeline stage forwards activations of size $2S_p h b$ bytes, while the decode phase transfers $2h b$ bytes per generated token.

The number of communication links equals $p-1$, as the first pipeline rank receives no input and the final rank produces no intermediate output. The total communication volume for pure pipeline parallelism is:

\begin{equation}
V_{pp} = (p-1) \times 2 \times (S_p + S_d - 1) \times h \times b
\end{equation}

\subsection{Hybrid Parallelism}

Hybrid parallelism combines tensor and pipeline strategies to enable efficient scaling across multiple nodes while maintaining computational efficiency within nodes. This approach introduces additional communication requirements, as received activations must be redistributed among tensor-parallel workers within each pipeline stage through Allgather operations.

The total communication volume for hybrid parallelism comprises four distinct components:

\begin{equation}
V_{hybrid} = V_{allreduce} + V_{allgather} + V_{gather} + V_{p2p}
\end{equation}

where each component is defined as:

\begin{align}
V_{allreduce} &= \frac{2L}{p} \times (S_p + S_d - 1) \times h \times b \times 2\left(\frac{t-1}{t}\right) \\
V_{allgather} &= 2(p-1) \times (S_p + S_d - 1) \times h \times b \times \left(\frac{t-1}{t}\right) \\
V_{gather} &= S_d \times \frac{v}{t} \times b \\
V_{p2p} &= (p-1) \times 2 \times (S_p + S_d - 1) \times \frac{h}{t} \times b
\end{align}

The Allreduce volume is reduced by a factor of $p$ due to layer distribution across pipeline stages, while Allgather operations enable activation redistribution within tensor-parallel groups. For the initial pipeline rank, an additional embedding layer contribution of $(S_p + S_d - 1) \times h \times b$ bytes applies to the Allreduce volume.

\section{Experiment Setup}
\label{sec:setup}

\begin{table}[ht]
    \centering
    \begin{tabular}{c|c}
    \toprule
        CPU & Intel Xeon Platinum 8470 (52 cores, 2 GHz) \\
        GPU & 4 × NVIDIA H100 (94 GB HBM2e with NVLink) \\
        Interconnect & InfiniBand NDR400 (4 NICs/Node) \\
        PyTorch Version & 2.6 \\
        vLLM Version & 0.8.5.post1 \\
        NCCL Version & 2.21.5 \\
    \bottomrule
    \end{tabular}
    \vspace{2ex}
    \caption{Experimental platform specifications.}
    \label{tab:cardinal}
\end{table}

All experiments were conducted on the OSC Cardinal supercomputer, with hardware and software specifications detailed in Table~\ref{tab:cardinal}. The compute topology of a Cardinal node features high-bandwidth NVLink connectivity between GPUs and InfiniBand networking for inter-node communication.

\subsection{Software Configuration}

To ensure consistent communication behavior across experiments, we configured vLLM with several key modifications. First, we disabled vLLM's custom allreduce implementation, directing all collective operations to the system NCCL library for standardized communication patterns. Second, we disabled PyTorch compilation to minimize performance variance across different parallelism configurations. Third, we enforced usage of vLLM's stable V0 engine rather than the actively developed V1 engine to maintain experimental consistency.

\subsection{Experimental Methodology}

Our evaluation focuses on single-request inference scenarios to isolate communication patterns from batching effects. Server metrics are extracted through RESTful API calls to the vLLM server endpoint, while detailed profiling data is collected using vLLM's integrated PyTorch profiler triggered via server requests. To ensure accurate communication measurements, we exclude rank-0 profiles from analysis to eliminate server initialization overhead and focus solely on inference-time communication patterns.

For model consistency, we utilize Llama model variants from Hugging Face, ensuring uniform transformer architecture across different parameter scales. This standardization enables direct comparison of communication characteristics across varying model sizes and parallelism configurations.

\section{Performance Characterization}
\label{sec:results}

This section presents experimental results and their integration with our analytical models to derive meaningful insights into distributed LLM inference communication patterns.

\subsection{Message Size and Frequency}

We extract communication operation counts and kernel calls from PyTorch profiler traces and validate them against our theoretical models across different parallelism strategies.

\textbf{Tensor Parallelism}: We collected profiling data for Llama-3.1-8B with $S_p = S_d = 128$ tokens, as presented in Table~\ref{tab:tp-count}. Our results demonstrate that varying TP degree does not affect Allreduce operation counts or message sizes, as counts depend solely on the number of transformer layers and decoding steps, while message sizes are determined by sequence length and hidden dimension. Notably, Gather message sizes scale inversely with TP workers since each worker gathers a partitioned slice of the vocabulary logits ($v$/$t$). The profiling reveals excellent alignment between our theoretical predictions and empirical measurements for both operation counts and message dimensions. We extend this analysis to Llama-3.2-3B and Llama-2-13B models in Table~\ref{tab:tp-model-count} to examine how communication patterns scale across different model architectures.

\begin{table*}[htbp!]
\centering
\begin{tabularx}{\textwidth}{c|c|XXX|XXX}
\toprule
\multirow{2}{*}{\textbf{Model}} & \multirow{2}{*}{\textbf{TP Size}} & \multicolumn{3}{c|}{\textbf{Prefill Stage}} & \multicolumn{3}{c}{\textbf{Decode Stage}} \\
\cmidrule(lr){3-5} \cmidrule(lr){6-8}
& & \textbf{Collective} & \textbf{Count} & \textbf{Shape} & \textbf{Collective} & \textbf{Count} & \textbf{Shape} \\
\midrule
\multirow{4}{*}{\parbox{1.5cm}{Llama-3.1-8B\\$S_p=128$\\$S_d=128$}} 
& \multirow{2}{*}{2} & Allreduce & 65 & [128,4096] & Allreduce & 8255 & [1,4096] \\
& & Gather & 1 & [64128] & Gather & 127 & [64128] \\
\cmidrule(lr){2-8}
& \multirow{2}{*}{4} & Allreduce & 65 & [128,4096] & Allreduce & 8255 & [1, 4096] \\
& & Gather & 1 & [32064] & Gather & 127 & [32064] \\
\bottomrule
\end{tabularx}
\vspace{2ex}
\caption{Message size and frequency breakdown for intra-node TP using Llama-3.1-8B}
\label{tab:tp-count}
\end{table*}

\begin{figure*}[htbp]
    \centering
    \begin{minipage}[b]{0.32\textwidth}
        \centering
        \includegraphics[width=\textwidth]{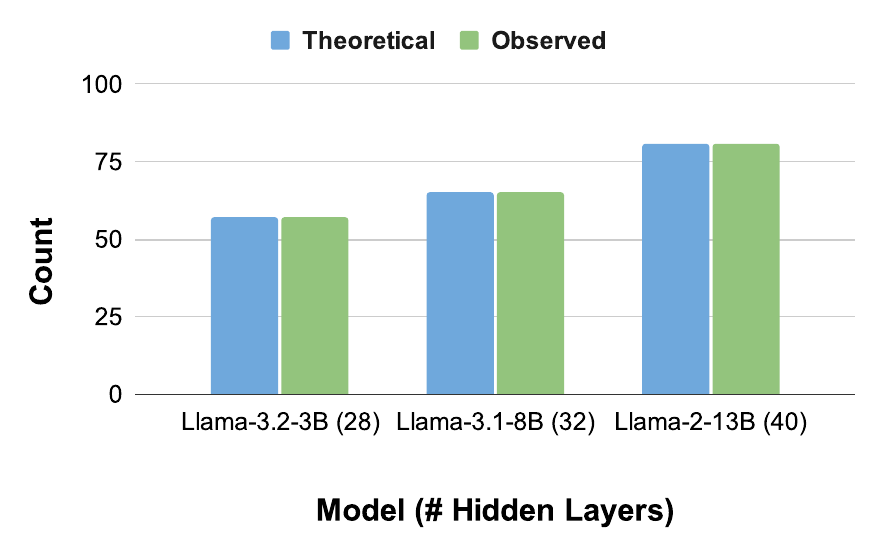}
        \caption*{(a) Prefill Allreduce Count}
        \label{fig:val-prefill-tp}
    \end{minipage}
    \hfill
    \begin{minipage}[b]{0.32\textwidth}
        \centering
        \includegraphics[width=\textwidth]{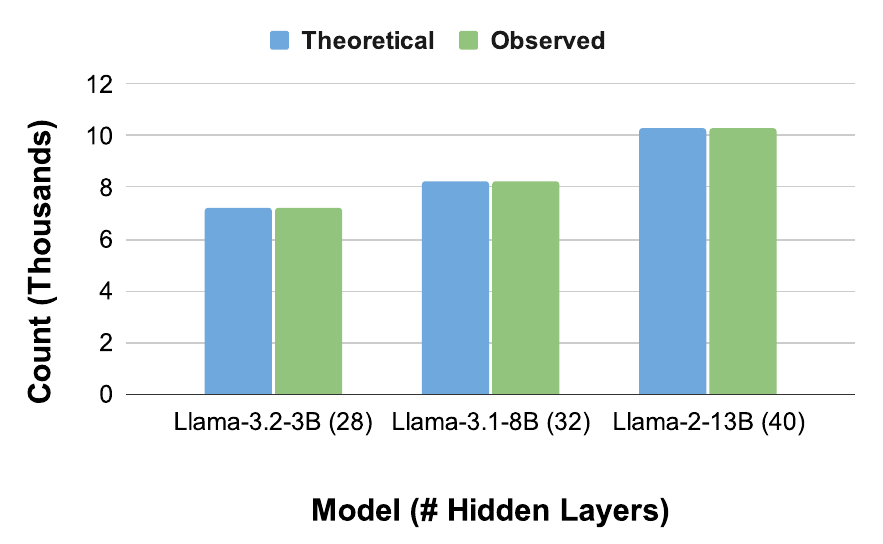}
        \caption*{(b) Decode Allreduce Count}
        \label{fig:val-decode-tp}
    \end{minipage}
    \hfill
    \begin{minipage}[b]{0.32\textwidth}
        \centering
        \includegraphics[width=\textwidth]{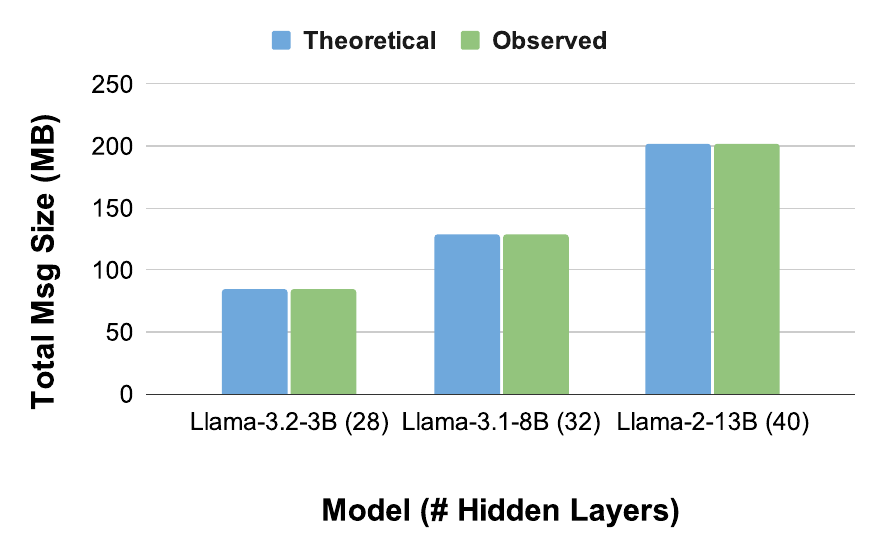}
        \caption*{(c) E2E Allreduce Total Message Size}
        \label{fig:val-e2e-tp}
    \end{minipage}
    \caption{Tensor Parallelism: Validation with observed data (Count \& Total Message Size), TP=4, across models}
    \label{fig:val-tp}
\end{figure*}

\begin{table*}[ht!]
\centering
\begin{tabular}{l|cc|cc|cc}
\toprule
& \multicolumn{2}{c|}{\textbf{Llama-3.2-3B}} & \multicolumn{2}{c|}{\textbf{Llama-3.1-8B}} & \multicolumn{2}{c}{\textbf{Llama-2-13B}} \\
\midrule
Message Size (bytes) & 786432 & 6144 & 1048576 & 8192 & 1310720 & 10240 \\
Count & 57 & 7239 & 65 & 8255 & 81 & 10287 \\
\bottomrule
\end{tabular}
\vspace{2ex}
\caption{Allreduce message size and count comparison across models for end-to-end inference}
\label{tab:tp-model-count}
\end{table*}

\textbf{Pipeline Parallelism}: In pipeline parallelism, transformer layers are distributed across GPU workers, requiring point-to-point communication to transfer intermediate activations between pipeline stages. Table~\ref{tab:pp-count} presents profiling results for Llama-3.1-8B with PP=2 and PP=4 configurations. The communication counts follow the pattern $(p-1) \times 2 \times KV_{factor}$, where the factor of 2 accounts for separate transmission of key and value tensors. Our empirical results align closely with theoretical predictions, confirming that point-to-point communication volume scales proportionally with the number of pipeline links. The message sizes remain small and depend primarily on the model's hidden dimension, with most communication occurring during the decode stage due to the autoregressive nature of transformer inference.

Figures \ref{fig:val-tp} and \ref{fig:val-pp} provide visual validation of our analytical models for tensor parallelism and pipeline parallelism respectively, demonstrating agreement between theoretical predictions and empirical measurements across different model sizes and parallelism degrees.

\begin{table*}[htbp!]
\centering
\begin{tabularx}{\textwidth}{c|c|XXX|XXX}
\toprule
\multirow{2}{*}{\textbf{Model}} & \multirow{2}{*}{\textbf{PP Size}} & \multicolumn{3}{c|}{\textbf{Prefill Stage}} & \multicolumn{3}{c}{\textbf{Decode Stage}} \\
\cmidrule(lr){3-5} \cmidrule(lr){6-8}
& & \textbf{Operation} & \textbf{Count} & \textbf{Shape} & \textbf{Operation} & \textbf{Count} & \textbf{Shape} \\
\midrule
\multirow{4}{*}{\parbox{1.5cm}{Llama-3.1-8B\\$S_p=128$\\$S_d=128$}} 
& \multirow{2}{*}{2} & Send & 2 & [128,4096] & Send & 254 & [1,4096] \\
& & Recv & 2 & [128,4096] & Recv & 254 & [1,4096] \\
\cmidrule(lr){2-8}
& \multirow{2}{*}{4} & Send & 6 & [128,4096] & Send & 762 & [1,4096] \\
& & Recv & 6 & [128,4096] & Recv & 762 & [1,4096] \\
\bottomrule
\end{tabularx}
\vspace{2ex}
\caption{Message size and frequency breakdown for pipeline parallelism}
\label{tab:pp-count}
\end{table*}

\begin{figure*}[htbp]
    \centering
    \begin{minipage}[b]{0.48\textwidth}
        \centering
        \includegraphics[width=\textwidth]{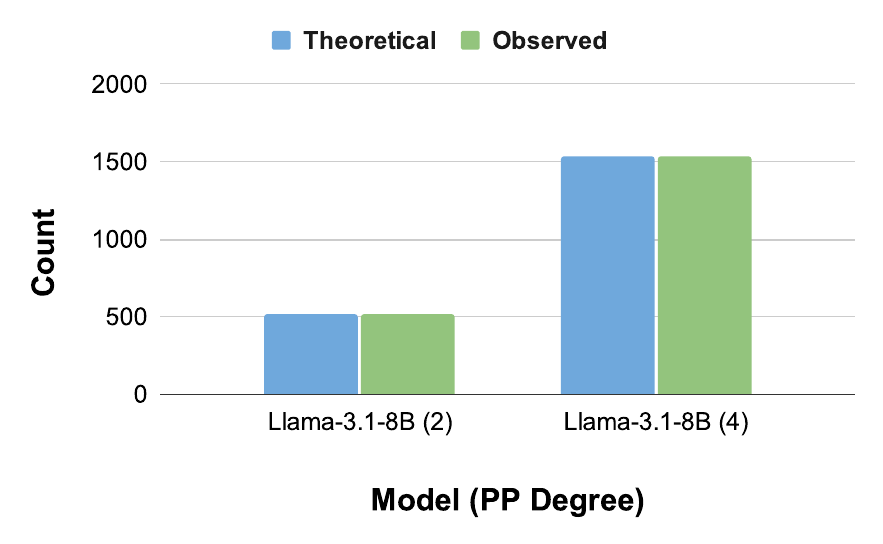}
        \caption*{(a) E2E point-to-point Count}
    \end{minipage}
    \hfill
    \begin{minipage}[b]{0.48\textwidth}
        \centering
        \includegraphics[width=\textwidth]{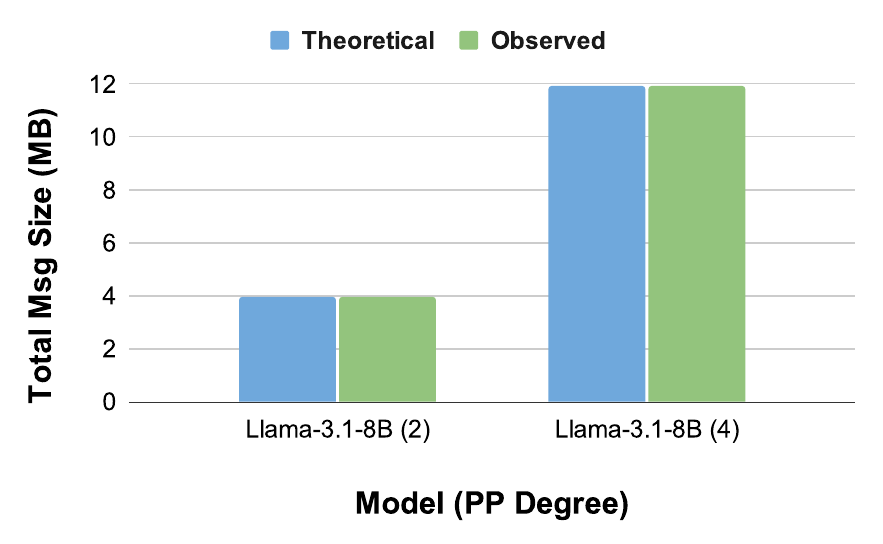}
        \caption*{(b) E2E point-to-point Total Message Size}
    \end{minipage}
    \caption{Pipeline Parallelism: Validation with observed data (Count \& Total Message Size), Llama-3.1-8B, across PP Degrees}
    \label{fig:val-pp}
\end{figure*}

\textbf{Hybrid Parallelism}: Combining tensor and pipeline parallelism introduces a more complex communication pattern involving four distinct operation types: Allreduce, Allgather, point-to-point transfers, and Gather operations. Table~\ref{tab:hybrid-count} presents results for TP=2, PP=2 configuration using Llama-3.1-8B. With 32 transformer layers distributed across 2 pipeline stages, each stage performs $(2L/p) + 1 = (2 \times 32/2) + 1 = 33$ Allreduce operations during prefill, where the additional operation stems from the parallel vocabulary embedding layer. The Allgather operations facilitate redistribution of received activations among tensor-parallel workers within each pipeline stage, while point-to-point transfers handle inter-stage communication with tensor dimensions adjusted for the TP degree ([128,2048] = [$S_p$, $h$/$t$]). The profiling data validates our analytical model, confirming that Allreduce operations constitute the majority of communication calls in hybrid configurations.

\begin{table*}[htbp!]
\centering
\begin{tabularx}{\textwidth}{l|l|XXX|XXX}
\toprule
\multirow{2}{*}{\textbf{Model}} & \multirow{2}{*}{\textbf{TP×PP}} & \multicolumn{3}{c|}{\textbf{Prefill Stage}} & \multicolumn{3}{c}{\textbf{Decode Stage}} \\
\cmidrule(lr){3-5} \cmidrule(lr){6-8}
& & \textbf{Operation} & \textbf{Count} & \textbf{Shape} & \textbf{Operation} & \textbf{Count} & \textbf{Shape} \\
\midrule
\multirow{4}{*}{\parbox{2.2cm}{Llama-3.1-8B\\$S_p=128$\\$S_d=128$}} 
& \multirow{4}{*}{2×2} & Allreduce & 33 & [128,4096] & Allreduce & 4191 & [1,4096] \\
& & Gather & 1 & [64128] & Gather & 127 & [64128] \\
& & Allgather & 2 & [128,4096] & Allgather & 254 & [1,4096] \\
& & Send/Recv & 2 & [128,2048] & Send/Recv & 254 & [1,2048] \\
\bottomrule
\end{tabularx}
\vspace{2ex}
\caption{Message size and frequency breakdown for hybrid parallelism (TP×PP) using Llama-3.1-8B}
\label{tab:hybrid-count}
\end{table*}

\textbf{Key Takeaways}: Our profiling analysis reveals three critical insights: (1) Communication operations exhibit moderate message sizes with high frequency, particularly during the decode stage where operations scale proportionally with sequence length; (2) The decode stage dominates communication volume due to the autoregressive nature of transformer inference, generating 127× more operations than prefill for $S_d = 128$; (3) Allreduce operations constitute the majority of collective communications in tensor parallelism and hybrid configurations, while pipeline parallelism relies primarily on point-to-point transfers. These findings emphasize the importance of optimizing high-frequency, moderate-sized collective operations when co-designing communication libraries for distributed LLM inference workloads.

\subsection{Communication Volume Breakdown}

\textbf{Parallelism Strategy Comparison}: We analyze total communication volume across different parallelism configurations using three representative models (Llama-3.2-3B, Llama-3.1-8B, Llama-2-13B) to understand how communication overhead varies with parallelism strategies under identical hardware constraints. Communication volume is calculated by multiplying total message size by appropriate correction factors: $2 \times \frac{d-1}{d}$ for Allreduce operations, $\frac{d-1}{d}$ for Allgather operations, and $1$ for point-to-point and Gather operations, where $d$ represents the number of participating workers \cite{nccl-test-eq}.

\begin{figure}[htbp]
    \centering
    \includegraphics[width=\columnwidth]{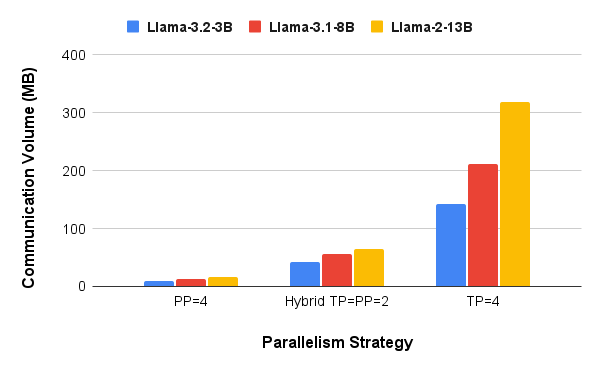}
    \caption{Communication volume comparison across parallelism strategies for LLM inference with $S_p$ = $S_d$ = 128 tokens using FP16/BF16 precision.}
    \label{fig:comm_volume_parallelism}
\end{figure}

Figure~\ref{fig:comm_volume_parallelism} reveals distinct communication characteristics across parallelism strategies. Pipeline parallelism (PP=4) demonstrates the lowest communication volume, achieving efficient scaling through minimal point-to-point transfers between pipeline stages. Conversely, tensor parallelism (TP=4) exhibits the highest communication overhead, primarily due to frequent Allreduce operations—two per transformer layer during decode—that scale with both model depth and sequence length. Hybrid parallelism (TP=2, PP=2) achieves a balanced middle ground , where layer distribution across pipeline stages significantly reduces Allreduce frequency while introducing manageable point-to-point communication overhead. The communication volume scales consistently with model size, reflecting the direct relationship between hidden dimensions, layer count, and message sizes across all parallelism strategies.

\textbf{Decode Sequence-Length Scaling}: We examine communication volume scaling with decode sequence length ($S_d$), a critical factor for real-world applications requiring long-form generation. This analysis uses fixed hardware resources (4 GPUs on a single node) organized into different parallelism configurations to isolate the impact of sequence length on communication patterns.

\begin{figure*}[htbp]
    \centering
    \begin{minipage}[b]{0.32\textwidth}
        \centering
        \includegraphics[width=\textwidth]{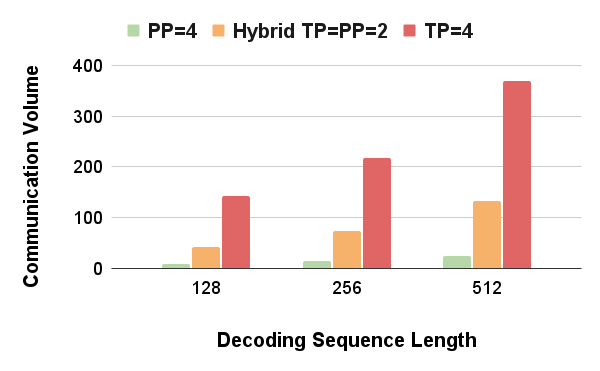}
        \caption*{(a) Llama-3.2-3B}
        \label{fig:seq-len-3B}
    \end{minipage}
    \hfill
    \begin{minipage}[b]{0.32\textwidth}
        \centering
        \includegraphics[width=\textwidth]{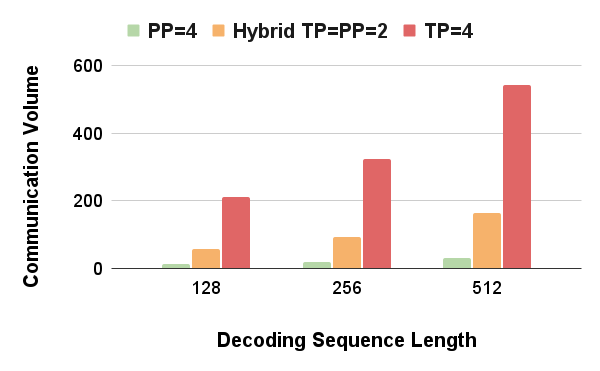}
        \caption*{(b) Llama-3.1-8B}
        \label{fig:seq-len-8B}
    \end{minipage}
    \hfill
    \begin{minipage}[b]{0.32\textwidth}
        \centering
        \includegraphics[width=\textwidth]{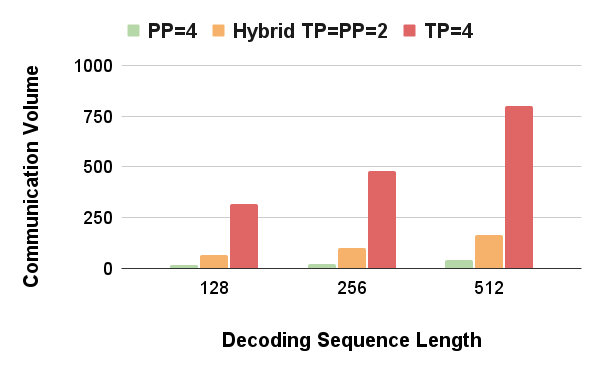}
        \caption*{(c) Llama-2-13B}
        \label{fig:seq-len-13B}
    \end{minipage}
    \caption{Communication volume scaling with decode sequence length across parallelism strategies for LLM inference with $S_p = 128$ tokens using FP16/BF16 precision.}
    \label{fig:seq-len-scale}
\end{figure*}

Figure~\ref{fig:seq-len-scale} demonstrates predictable yet non-linear scaling behavior across all models and parallelism strategies. Several key patterns emerge: (1) Communication volume increases sub-linearly with decode length due to the fixed prefill term $S_p$ in our analytical formulas—as $S_d$ increases from 128 to 512 tokens (4× growth), communication volume increases by approximately 2.5×, consistent with the $(S_p + S_d - 1)$ scaling relationship; (2) Pipeline parallelism maintains the most predictable and lowest absolute communication volume, with clean linear scaling governed by point-to-point transfer requirements; (3) Tensor parallelism exhibits dramatic volume growth, becoming prohibitive for long sequences due to both Allreduce scaling with $(S_p + S_d - 1)$ and additional Gather operations scaling directly with $S_d$; (4) Hybrid parallelism provides reasonable scaling behavior but approaches communication-bound regimes for the longest sequences.

The scaling analysis validates our theoretical predictions, with observed growth factors matching analytical expectations: 1.50× for 128→256 tokens and 1.67× for 256→512 tokens. This sub-linear scaling occurs because the fixed prefill cost ($S_p = 128$) increasingly dilutes the relative impact of decode sequence growth, though absolute volumes still reach problematic levels for tensor parallelism at long sequence lengths.

\textbf{Key Takeaways}: Our analysis reveals fundamental trade-offs in parallelism selection for distributed LLM inference. Tensor parallelism, while offering superior computational parallelization, incurs substantial communication overhead that scales unfavorably with both model size and sequence length. Pipeline parallelism maintains consistent, minimal communication pressure on network infrastructure, making it ideal for bandwidth-constrained environments and long-sequence applications. Hybrid parallelism offers a viable compromise for moderate workloads but requires careful configuration to avoid the communication penalties observed in poorly balanced arrangements. These findings provide crucial guidance for deployment strategy selection based on infrastructure capabilities and application requirements.

\subsection{Service Level Objective Evaluation}

We evaluate the impact of different parallelism strategies on critical service level objectives (SLOs) for distributed LLM inference, focusing on End-to-End latency, Time-to-First-Token (TTFT), and Time-Ter-Output-Token (TPOT) metrics that directly affect user experience in production deployments.

\textbf{Tensor Parallelism Scaling}: We investigate the effect of increasing tensor parallelism degree on SLO metrics using Llama-3.2-3B with $S_p = S_d = 128$ tokens. Our evaluation covers TP=2, 4, and 8, where TP=8 spans two nodes with 4 GPUs each, allowing us to examine both intra-node and inter-node scaling behavior.

\begin{figure*}[htbp]
    \centering
    \begin{minipage}[b]{0.32\textwidth}
        \centering
        \includegraphics[width=\textwidth]{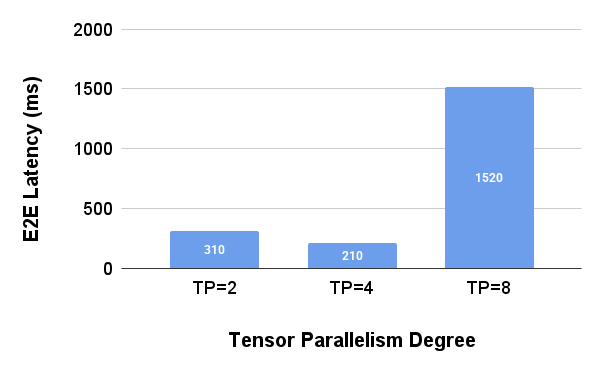}
        \caption*{(a) End-to-end Latency}
        \label{fig:tp-e2e}
    \end{minipage}
    \hfill
    \begin{minipage}[b]{0.32\textwidth}
        \centering
        \includegraphics[width=\textwidth]{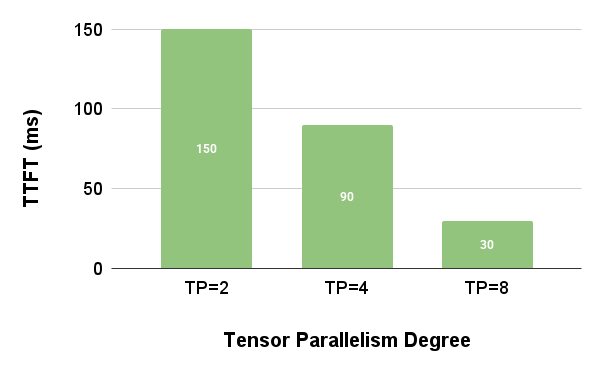}
        \caption*{(b) Time-to-first-token}
        \label{fig:tp-ttft}
    \end{minipage}
    \hfill
    \begin{minipage}[b]{0.32\textwidth}
        \centering
        \includegraphics[width=\textwidth]{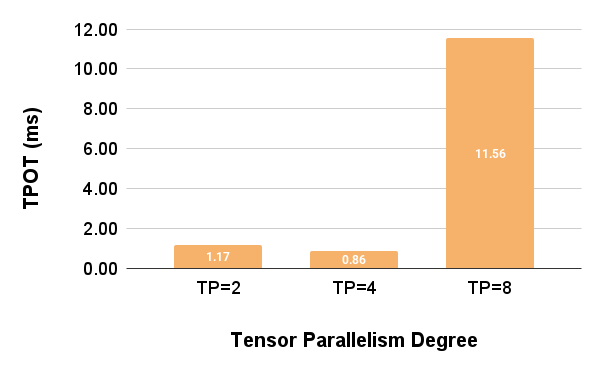}
        \caption*{(c) Time-per-output-token}
        \label{fig:tp-tpot}
    \end{minipage}
    \caption{Llama-3.2-3B SLO metrics comparison across tensor parallelism degrees for LLM inference with $S_p = S_d = 128$ tokens using FP16/BF16 precision.}
    \label{fig:tp-slo}
\end{figure*}

Figure~\ref{fig:tp-slo} reveals distinct scaling patterns across SLO metrics. Scaling from TP=2 to TP=4 yields substantial improvements across all metrics: end-to-end latency decreases from 310ms to 210ms, TTFT improves from 150ms to 90ms, and TPOT reduces from 1.17ms to approximately 0.86ms. This improvement stems from effective computational workload distribution across GPUs within a single node, where high-bandwidth NVLink interconnects enable efficient Allreduce operations while the increased parallelization significantly reduces per-GPU computational load.

However, scaling to TP=8 across two nodes produces mixed results. While TTFT continues to improve (from 90ms to 30ms), demonstrating that the prefill stage—being compute-bound—benefits from maximum parallelization regardless of communication overhead, both end-to-end latency and TPOT degrade significantly (latency increases to 1520ms, TPOT rises to 11.56ms). This degradation occurs because the decode stage becomes communication-bound when using lower-bandwidth inter-node networks, where the increased Allreduce frequency overwhelms the computational benefits of additional parallelization.

\textbf{Pipeline Parallelism Scaling}: We examine the impact of increasing pipeline parallelism degree on SLO metrics using Llama-3.2-3B with fixed prefill and decode sequence-lengths of 128 tokens. Our evaluation covers PP=2, 4, and 8, where PP=8 spans two nodes to assess inter-node pipeline performance characteristics.

Figure \ref{fig:pp-slo} reveals scaling patterns that highlight pipeline parallelism's limitations for latency-sensitive applications. Scaling from PP=2 to PP=4 shows significant performance degradation: end-to-end latency increases from 0.69s to 1.36s, TTFT rises from 430ms to 1110ms, while TPOT remains relatively stable at approximately 2ms. This degradation stems from increased pipeline depth creating longer dependency chains, where each stage must wait for upstream computations to complete before processing can begin.

The performance degradation becomes severe at PP=8 across two nodes, with end-to-end latency reaching 4.98s (6× worse than PP=2), TTFT climbing to 2520ms (5× degradation), and TPOT spiking to 19.22ms (6× increase). This dramatic performance loss occurs because pipeline parallelism inherently serializes the inference process—each stage must complete its computation before the next stage can begin, creating cumulative latency that scales with pipeline depth. Additionally, inter-node communication introduces substantial overhead for the frequent point-to-point transfers required between pipeline stages.

These results demonstrate that while pipeline parallelism minimizes communication volume (as shown in Section~\ref{sec:analysis}), it fundamentally trades latency for memory efficiency and communication reduction. The sequential nature of pipeline processing makes it unsuitable for interactive applications requiring low response times, despite its advantages for long-sequence generation and bandwidth-constrained environments where communication volume, rather than latency, is the primary constraint.

\begin{figure*}[htbp]
    \centering
    \begin{minipage}[b]{0.32\textwidth}
        \centering
        \includegraphics[width=\textwidth]{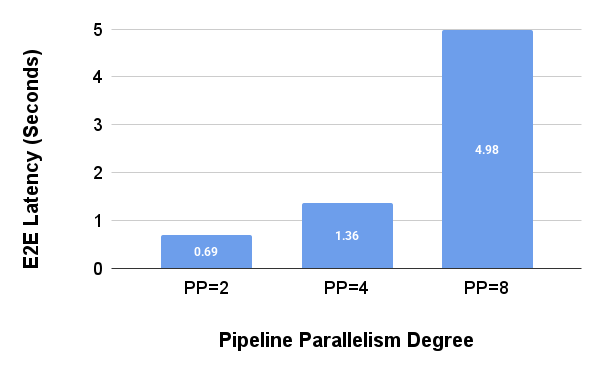}
        \caption*{(a) End-to-end Latency}
        \label{fig:pp-e2e}
    \end{minipage}
    \hfill
    \begin{minipage}[b]{0.32\textwidth}
        \centering
        \includegraphics[width=\textwidth]{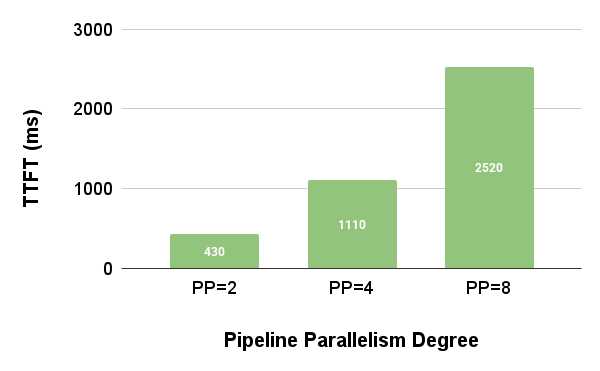}
        \caption*{(b) Time-to-first-token}
        \label{fig:pp-ttft}
    \end{minipage}
    \hfill
    \begin{minipage}[b]{0.32\textwidth}
        \centering
        \includegraphics[width=\textwidth]{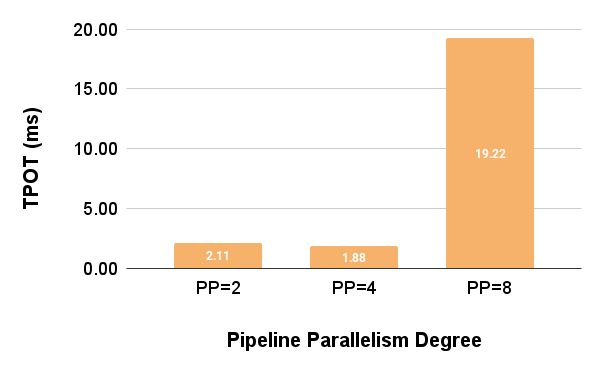}
        \caption*{(c) Time-per-output-token}
        \label{fig:pp-tpot}
    \end{minipage}
    \caption{Llama-3.2-3B SLO metrics comparison across pipeline parallelism degrees for LLM inference with $S_p = S_d = 128$ tokens using FP16/BF16 precision.}
    \label{fig:pp-slo}
\end{figure*}

\textbf{Hybrid Parallelism Strategy Comparison}: We evaluate various hybrid parallelism configurations using Llama-2-13B deployed across 8 GPUs on two nodes, examining how different TP/PP combinations affect real-time serving performance.

\begin{figure*}[htbp]
    \centering
    \begin{minipage}[b]{0.32\textwidth}
        \centering
        \includegraphics[width=\textwidth]{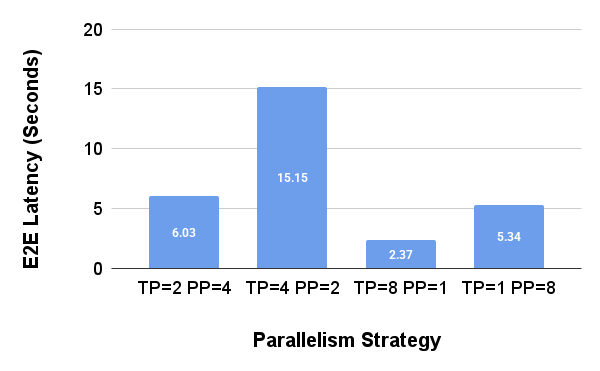}
        \caption*{(a) End-to-end Latency}
        \label{fig:hybrid-e2e}
    \end{minipage}
    \hfill
    \begin{minipage}[b]{0.32\textwidth}
        \centering
        \includegraphics[width=\textwidth]{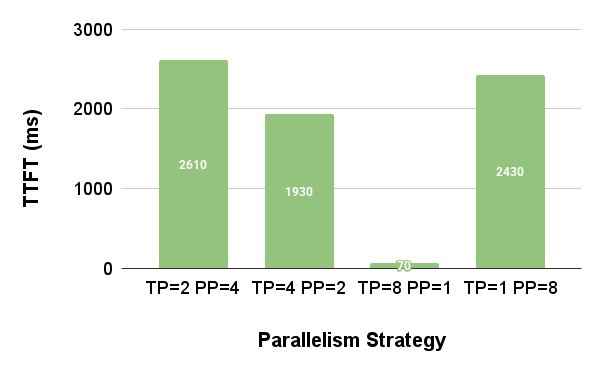}
        \caption*{(b) Time-to-first-token}
        \label{fig:hybrid-ttft}
    \end{minipage}
    \hfill
    \begin{minipage}[b]{0.32\textwidth}
        \centering
        \includegraphics[width=\textwidth]{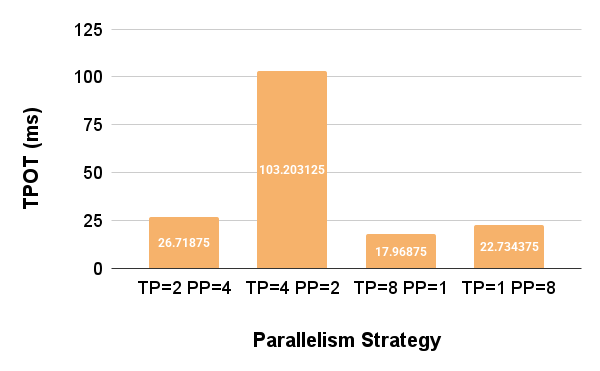}
        \caption*{(c) Time-per-output-token}
        \label{fig:hybrid-tpot}
    \end{minipage}
    \caption{Llama-2-13B SLO metrics comparison across hybrid parallelism strategies for LLM inference with $S_p = S_d = 128$ tokens using FP16/BF16 precision.}
    \label{fig:hybrid-slo}
\end{figure*}

Figure~\ref{fig:hybrid-slo} demonstrates the critical importance of parallelism strategy selection for large model deployments. Pure tensor parallelism (TP=8 PP=1) achieves exceptional performance across all metrics: 2.37s end-to-end latency, 70ms TTFT, and 18ms TPOT, representing the optimal configuration for this workload. The superior TTFT performance (70ms vs 1930-2610ms for other configurations) reflects tensor parallelism's ability to parallelize prefill computation across all available GPUs simultaneously.

Pure pipeline parallelism (TP=1 PP=8) delivers moderate performance with 2430ms TTFT and reasonable end-to-end latency, benefiting from minimal communication overhead but suffering from sequential processing limitations during prefill. The balanced hybrid configuration (TP=2 PP=4) provides intermediate performance across all metrics, while the unbalanced configuration (TP=4 PP=2) performs catastrophically with 15.15s end-to-end latency and 103ms TPOT, illustrating the severe penalties of poorly configured hybrid strategies.

\textbf{Key Takeaways}: Based on our comprehensive SLO analysis, we recommend pure tensor parallelism (TP=8) for interactive applications requiring ultra-low latency with short sequences, where TTFT is critical and high-bandwidth interconnects are available, as computational parallelization benefits overwhelm communication overhead for short workloads. Conversely, pure pipeline parallelism (PP=8) suits long-form generation tasks, memory-constrained environments, or communication-limited infrastructures, maintaining predictable performance while minimizing communication volume. Hybrid parallelism provides a viable middle ground for balanced production workloads with moderate sequence lengths, but requires careful configuration with balanced arrangements (e.g., TP=2 PP=4) to avoid the catastrophic performance penalties observed in unbalanced configurations (TP=4 PP=2). The fundamental insight is that while computational parallelization can overwhelm communication overhead for short sequences, this advantage diminishes with longer sequences and inter-node deployments, where communication becomes the primary performance bottleneck, necessitating a shift toward communication-efficient strategies.

\section{Related Work}
\label{sec:related}
Research into optimizing Large Language Model (LLM) inference is extensive, focusing on systems design, performance characterization, and parallelism strategies. While many studies address overall efficiency, detailed systematic analysis of communication behavior across diverse parallelism strategies in modern inference frameworks remains underexplored.
Several works have characterized Deep Neural Network (DNN) performance on HPC systems \cite{Awan-DNN-Char, Jain-DNN-Char-PyTorch}, with some evaluating specific aspects like CUDA-aware MPI performance \cite{Awan-CUDA-Aware-MPI-Char}. More recently, LLM-specific system performance has been analyzed. For instance, Yin et al. \cite{LLM-Arch} investigate how different LLM architectures perform on supercomputers, while Hu et al. \cite{LLM-datacenter} explore LLM impacts on datacenter hardware, noting communication as a factor in performance degradation but without providing in-depth communication analysis.
Anthony et al. \cite{anthony2024demystifying} provide valuable communication characterization for distributed transformer models during training. However, our work specifically focuses on the distinct prefill and decode phases of inference, offering analytical models for communication volume within a production-grade inference server (vLLM \cite{kwon2023efficient}).
The development of LLM inference serving systems like vLLM \cite{kwon2023efficient}, Text Generation Inference, and DeepSpeed-Inference \cite{aminabadi2022deepspeed} has brought significant improvements through techniques like PagedAttention \cite{kwon2023efficient} and optimized kernels. Zhong et al. \cite{zhong2024distserve} highlight the importance of disaggregating prefill and decoding stages for improved efficiency. While these systems improve overall performance, our contribution lies in analyzing their communication patterns under Tensor Parallelism (TP), Pipeline Parallelism (PP), and hybrid approaches, providing empirically validated analytical models for communication volume (Section~\ref{sec:analysis}) that are not typically detailed in these system papers.
Systems like Orca \cite{yu2022orca} and Alpa \cite{zheng2022alpa} explore co-optimization of parallelism strategies and batching for LLM inference, often focusing on automated configuration. Pope et al. \cite{pope2022efficiently} discuss scaling transformer inference efficiently, addressing memory and compute considerations. Our study complements these works by providing granular breakdowns of communication statistics—including message sizes, frequencies, and collective operation types—for explicit TP, PP, and hybrid configurations (Section~\ref{sec:results}).
This detailed characterization, combined with insights into how these patterns affect end-user service level objectives across different model sizes and sequence lengths, offers foundational understanding for selecting and optimizing parallelism configurations. Unlike prior work \cite{LLM_Scale_out_Char} that characterized LLM performance at scale with emphasis on network utilization, our work directly models and measures communication volume and its specific behavior under varied parallelism schemes and inference stages, linking these directly to user-facing metrics.
\section{Future Work}
\label{sec:future}
Our analytical models provide a framework for predicting communication patterns in distributed LLM inference that presents opportunities for extension by the research community. The theoretical insights could be leveraged to develop automated parallelism selection tools that dynamically choose optimal configurations based on infrastructure characteristics and workload requirements, bridging the gap between analysis and practical deployment guidance. The current platform-specific validation would benefit from characterization across diverse hardware architectures including AMD and Intel GPUs with different interconnect topologies to establish broader generalizability. Similarly, extending this characterization methodology beyond vLLM to frameworks such as SGLang \cite{zheng2024sglangefficientexecutionstructured}, TensorRT-LLM \cite{nvidiaWelcomeTensorRTLLMx2019s}, and DeepSpeed-Inference \cite{aminabadi2022deepspeed} would enable framework-agnostic communication modeling and reveal engine-specific optimization opportunities. The communication patterns of emerging paradigms including mixture-of-experts models, speculative decoding, and unified memory architectures represent fertile ground for investigation using similar analytical approaches. 

\section{Conclusions}
\label{sec:conclusions}
We have presented a comprehensive characterization of communication patterns in distributed Large Language Model inference across multiple parallelism strategies. This has been accomplished by combining rigorous analytical models with extensive experimental validation using state-of-the-art inference frameworks and precise profiling of communication operations.
Our key findings reveal that communication characteristics vary dramatically across parallelism strategies. Tensor parallelism exhibits the highest communication volume but optimal latency for short sequences, pipeline parallelism demonstrates minimal communication overhead at the cost of increased latency, and hybrid parallelism offers a balanced compromise requiring careful configuration. The decode stage dominates communication volume due to autoregressive inference, with Allreduce operations constituting the majority of collective communications.
Our service level objective evaluation provides actionable deployment guidance on optimal parallelism layout given application scenarios. We support our insights with extensive analysis and scaling studies.
For future work, we plan to extend this characterization to emerging parallelism strategies such as expert parallelism and sequence parallelism. We also intend to evaluate these communication patterns on next-generation hardware architectures with advanced interconnects, including unified memory architectures with advanced offloading capabilities.
\section{Acknowledgements}
\label{sec:ack}

This research is supported in part by NSF grants \#2311830, \#2312927, \#2323116, \#2415201, \#2504944, and XRAC grant \#NCR-130002.

\bibliographystyle{IEEEtran}
\bibliography{inference.bib}

\end{document}